\documentclass[12pt]{iopart}

\usepackage{iopams}

\begin{document}

\title[Born-Infeld extension of Lovelock brane gravity]{Born-Infeld 
extension of Lovelock brane gravity}

\author{Miguel Cruz\dag and Efra\'\i n Rojas\ddag}

\address{\dag\ Departamento de F\'\i sica, Centro de Investigaci\'on
y de Estudios Avanzados del IPN, Apdo Postal 14-740, 07000
M\'exico DF, M\'exico}

\address{\ddag\ Departamento de F\'\i sica, Facultad de F\'\i sica e 
Inteligencia Artificial, Universidad Veracruzana, 91000 Xalapa, Veracruz, M\'exico}

\eads{\mailto{mcruz@fis.cinvestav.mx},\,\,\mailto{efrojas@uv.mx}}

\begin{abstract}
We present a Born-Infeld type theory to describe the evolution of 
$p$-branes propagating in an $N=(p+2)$-dimensional Minkowski spacetime. 
The expansion of the BI-type volume element gives rise to the $(p+1)$ 
Lovelock brane invariants associated with the worldvolume swept out by 
the brane. Contrary to the Lovelock theory in gravity, the number of 
Lovelock brane Lagrangians differs in this case, depending on the dimension
of the worldvolume as a consequence that we consider the embedding functions,
instead of the metric, as the field variables. This model depends on the 
intrinsic and the extrinsic geometries of the worldvolume and in consequence 
is a second-order theory as shown in the main text. A classically 
equivalent action is discussed and we comment on its Weyl invariance in any 
dimension which naturally requires the introduction of some auxiliary fields.  
\end{abstract}

%\submitto{\CQG}


%\maketitle

\date{\today}

%%%%%%%%%%%%%%%%%%%%%%%%%%%%%%%%%%%%%%%%%%%%%%%%%%%%%%%%%%%%%%%%%%

\section{Introduction}

%%%%%%%%%%%%%%%%%%%%%%%%%%%%%%%%%%%%%%%%%%%%%%%%%%%%%%%%%%%%%%%%%%

The usefulness of Born-Infeld structures has grown by leaps and bounds over the 
past few years because of their attractive geometric properties and capacity 
to implement some finite bounds in several physical theories~\cite{born,leigh,deser,gibbons,schuller,vollick,nieto,banados,tekin}.
These structures have found a natural place in the brane context as they may 
contain unknown dynamic symmetries, thus providing physical implications. 
Related constructions including higher order brane curvature terms but ignoring 
electromagnetism have also been considered~\cite{lindstrom,bib}. However, among the 
plethora of these approaches, no purely geometrical BI-like construction combining 
both the intrinsic and extrinsic geometries of a brane trajectory, leading to 
second-order equations, has been proposed. In this sense, it seems reasonable to 
ask whether exists a connection between BI structures and the Lovelock invariants
defined on the worldvolume.

The Lovelock brane invariants are second-order geometric terms constructed with
the extrinsic curvature of the worldvolume swept out by the $p$-brane that yield
second-order equations of motion. This fact is important because it ensures no 
propagation of extra degrees of freedom and this renders a given system free 
from many of the pathologies that plague higher order derivative theories. 
Interest in the Lovelock brane theory has attracted much attention recently, 
not only for its rich structure but also for its applications to describe cosmological phenomena~\cite{dgp,rham1,trodden2}. 

The purpose of this work is to provide a BI-type brane theory that describes 
the evolution of $p$-dimensional branes propagating in an $N=(p+2)$-dimensional 
Minkowski spacetime, which may be written in terms of the Lovelock brane invariants associated with the
co-dimension 1 worldvolume. The Lagrangian density of such a model is written 
close in spirit to the one developed in~\cite{deser}. This contains both contribution 
from the intrinsic and the extrinsic geometries through the first and second fundamental 
forms $g_{ab}$ and $K_{ab}$, respectively, inherited by the co-dimension 1 worldvolume. 
We find that the expansion of this alternative volume element casts out the Lovelock 
brane invariants for a given dimension $p$, which can be collected in a finite series.
To explore the structure of the Lovelock brane invariants, we exploit the 
so-called {\it conserved stress tensor}~\cite{noether}. This will permit us to 
know their mechanical content in an effortless way. 
In fact, this tool is nothing but the Noether current associated with the 
translation invariance of the theory. In addition, to complement our proposal 
we examine a classically equivalent action and we discuss its Weyl invariance 
through the introduction of an auxiliary metric and a scalar field. This may be 
useful in the construction of a Hamiltonian formulation for the theory 
thus avoiding the higher nonpolynomial dependence of the field variables.

This paper is organized as follows. The aim of Section 2 is to acquaint the reader
with the basic facts of the Lovelock theory for branes propagating in a flat 
background spacetime which will be used throughout the paper. We show that the 
Lovelock brane invariants can all be obtained by using antisymmetric products 
of the extrinsic curvature tensor and the Gauss-Codazzi integrability condition 
for surfaces. We obtain the general equations of motion in terms of the conserved 
tensors. In Section 3, we introduce a BI-type action containing the induced metric 
and the extrinsic curvature whose curvature expansion leads to the Lovelock brane 
Lagrangians for a given dimension $p$. We also discuss the Weyl invariance of a 
classically equivalent action. We conclude in Section 4 with some comments and we
discuss our results. Appendices A and B gather information about the standard Lovelock 
theory notation and some mathematical relations useful for expanding the BI-like 
structures.

%%%%%%%%%%%%%%%%%%%%%%%%%%%%%%%%%%%%%%%%%%%%%%%%%%%%%%%%%%%%%%%%%%%

\section{Lovelock brane theory}

%%%%%%%%%%%%%%%%%%%%%%%%%%%%%%%%%%%%%%%%%%%%%%%%%%%%%%%%%%%%%%%%%%

Consider a spacelike brane of dimension $p$ floating in a flat background spacetime
of dimension $N=(p+2)$ with metric $\eta_{\mu \nu}$ $(\mu,\nu =0,1,\ldots, p+1)$. 
The brane sweeps out an oriented hypersurface manifold of dimension $p+1$, known 
as worldvolume and denoted by $\Sigma$, described by the embedding functions $x^\mu 
= X^\mu(\xi^a)$ where $x^\mu$ are the local coordinates in the background spacetime 
and $\xi^a$ are the local coordinates for $\Sigma$ and $X^\mu$ are the embedding functions 
$(a,b= 0,1,\ldots,p)$. The only geometrically significant derivatives of $X^\mu$ are 
encoded in the induced metric tensor $g_{ab} = \eta_{\mu \nu} e^\mu{}_a e^\nu{}_b=: 
e_a \cdot e_b$ and the extrinsic curvature tensor $K_{ab} = - n \cdot \nabla_a e_b = 
K_{ba}$, where $e^\mu{}_a = \partial_a X^\mu$ are the tangent vectors to the worldvolume 
and $n^\mu$ denotes the normal vector to $\Sigma$ such that $n\cdot e_a = 0$, and 
$\nabla_a$ is the covariant derivative compatible with the induced metric $g_{ab}$.

For a $(p+1)$-dimensional worldvolume whose field variables are the embedding 
functions, the action
\begin{equation}
S[X] =  \int d^{p+1} \xi \, \sqrt{-g} \sum_{n=0} ^{p+1} \alpha_n\,L_n (g_{ab},
K_{ab}),
\label{eq:Lbaction}
\end{equation}
where
\begin{equation}
L_n (g_{ab}, K_{ab}) = \delta^{a_1 a_2 a_3 \cdots a_n} _{b_1 b_2 b_3 \cdots b_n}
 K^{b_1}{}_{a_1} K^{b_2}{}_{a_2} K^{b_3}{}_{a_3} \cdots K^{b_n}{}_{a_n},
\label{eq:lovelock-brane}
\end{equation}
ensures that the field equations of motion are of second order, as we will see below. 
Here, $\delta^{a_1 a_2 a_3 \ldots a_n} _{b_1 b_2 b_3 \ldots b_n}$ denotes the generalized
Kronecker delta (gKd) (see \ref{app:1} for details), $g= \textrm{det} (g_{ab})$ and 
$\alpha_n$ are the generic coefficients with appropriate dimensions. Furthermore, we 
set $L_0 = 1$. The action~(\ref{eq:Lbaction}) is a second-order derivative theory 
which is invariant under reparameterizations of the worldvolume. 
%By construction, these Lovelock brane terms 
%vanish for $n>p+1$ whereas the term with $n= p+1$ corresponds to a topological 
%invariant not contributing to the field equations. For example, for 
%the relativistic string case ($p=1$) the Lovelock term $L_2 = {\cal R}$ (see 
%below for details) corresponds to a topological invariant because of the Gauss-Bonnet 
%theorem. 
Since the independent variables are the embedding functions instead of the metric, 
we have a greater number of Lovelock brane terms contrary to the standard gravity 
case~\cite{lovelock}. By expanding out equation (\ref{eq:lovelock-brane}) in terms of 
minors we have
\begin{eqnarray} 
\fl
L_n =\left[ \delta^{a_1} _{b_1}\delta^{a_2 a_3 a_4 \cdots a_n} _{b_2 b_3 b_4 \cdots b_n}
- \delta^{a_1} _{b_2}\delta^{a_2 a_3 a_4 \cdots a_n} _{b_1 b_3 b_4 \cdots b_n}
+ \delta^{a_1} _{b_3}\delta^{a_2 a_3 a_4 \cdots a_n} _{b_1 b_2 b_4 \cdots b_n}
+ \cdots + (-1)^{n-1} \delta^{a_1} _{b_n}\delta^{a_2 a_3 a_4 \cdots a_n} _{b_1 b_2 b_3 
\cdots b_{n-1}}
\right] \times  
\nonumber
\\
 K^{b_1}{}_{a_1} K^{b_2}{}_{a_2} K^{b_3}{}_{a_3}\cdots K^{b_n}{}_{a_n} ,
\nonumber
\\
= KL_{n-1} - \left( n-1 \right)\, \delta^{a_2 a_3 a_4 \cdots a_n} _{b_2 b_3 b_4 
\cdots b_n} K^{b_2}{}_c K^c{}_{a_2} K^{b_3}{}_{a_3} \cdots K^{b_n}{}_{a_n},
\end{eqnarray}
where we have relabeled some indexes besides using the antisymmetric properties
of the gKd. The iterative expansion of the remaining gKd in the latter 
equation yields
\begin{equation}
L_n = \sum_{r=1} ^{n} \frac{(-1)^{r-1} (n-1)!}{(n-r)!}\,K_{ab} ^{r}\,
L_{n-r},
\label{eq:sum1}
\end{equation}
where $n=1,2,3,\ldots, p+1$ and we have introduced the useful short-hand notation 
$K_{ab} ^1 :=K^a{}_a, K_{ab} ^2 := K^a{}_b K^b{}_a, K_{ab} ^3 := K^a{}_b K^b{}_c 
K^c{}_a$ and so on. From this recursion formula, we can compute the first {\it Lovelock 
brane Lagrangians} 
\numparts
\begin{eqnarray}
L_0 &=& 1,
\label{eq:love1}
\\
L_1 &=& K,
\label{eq:love2}
\\
L_2 &=& K^{2} - K_{ab} ^2 = {\cal R},
\label{eq:love3}
\\
L_3 &=& K^{3} - 3 K K_{ab}^2 + 2 K_{ab} ^3 = -K_{ab} ^3 +
\frac{3}{2}K_{ab} ^2 - \frac{1}{2}K^3 - 3G_{ab}K^{ab},
\label{eq:love4}
\\
L_4 &=& K^4 - 6 K^2 K_{ab}^2 + 8 K K_{ab} ^3 + 3\left( K_{ab} ^2 \right)^2
- 6 K_{ab} ^4, \nonumber
\\
&=& {\cal R}^2 - 4 {\cal R}_{ab} {\cal R}^{ab} + {\cal R}_{abcd}
{\cal R}^{abcd},
\label{eq:love5}
\\
L_5 &=& K^5 -10 K^3 K_{ab} ^2 + 20 K^2 K_{ab} ^3 - 30 K K_{ab} ^4 +
15 K \left( K_{ab} ^2 \right)^2 - 20 K_{ab} ^2 K_{ab} ^3
\nonumber
\\
&+&  24 K_{ab} ^5,
\label{eq:love6}
\\
L_6 &=& K^6 - 15 K^4 K_{ab} ^2 + 40 K^3 K_{ab} ^3 - 90 K^2 K_{ab} ^4
+ 45 K^2 \left( K_{ab} ^2 \right)^2 
\nonumber
\\
&-& 120 K K_{ab} ^2 K_{ab} ^3 
+ 144 K K_{ab} ^5 + 90 K_{ab} ^2 K_{ab} ^4  - 15 \left( K_{ab} ^2 \right)^3 
+ 40 \left( K_{ab}^3 \right)^2
\nonumber
\\
&-&120 K_{ab} ^6, \nonumber
\\
&=& {\cal R}^3 - 12{\cal R} {\cal R}_{ab}{\cal R}^{ab} + 16 {\cal R}_{ab}
{\cal R}^a{}_c {\cal R}^{bc} + 24 {\cal R}_{ab} {\cal R}_{cd} {\cal R}^{acbd}
\nonumber
\\
&+& 3{\cal R} {\cal R}_{abcd}{\cal R}^{abcd} 
- 24 {\cal R}_{ab} {\cal R}^a{}_{cde} {\cal R}^{bcde} 
+ 2 {\cal R}_{abcd}{\cal R}^{ab}{}_{ef}{\cal R}^{cdef} 
\nonumber
\\
&-& 8 {\cal R}_{abcd} 
{\cal R}^a{}_e{} ^{c}{}_f {\cal R}^{bfde}.
\label{eq:love7}
%\\
%L_7 &=& K^7 - 21 K^5 K_{ab} ^2 + 70 K^4 K_{ab} ^3 - 210 K^3 K_{ab} ^4 + 105 K^3 
%\left(K_{ab} ^2\right)^2 
%\nonumber
%\\
%&-& 420 K^2 K_{ab} ^2 K_{ab} ^3  
%+ 504 K^2 K_{ab} ^5 - 840 K K_{ab}^6 + 630 K K_{ab} ^2 K_{ab} ^4 
%\nonumber
%\\
%&-& 105 K 
%\left( K_{ab} ^2 \right)^3 
%+ 280 K \left( K_{ab} ^3 \right)^2 - 420 K_{ab} ^3 K_{ab} ^4
%+ 210 K_{ab} ^3 \left( K_{ab} ^2 \right)^2 
%\nonumber
%\\
%&-& 504 K_{ab} ^5 K_{ab} ^2 + 720 K_{ab} ^7,
%\label{eq:love8}
\end{eqnarray}
\endnumparts
Here, we have used repeatedly the contracted Gauss-Codazzi integrability condition 
in a flat spacetime background, ${\cal R}_{abcd} = K_{ac} K_{bd} - K_{ad}K_{bc}$, where 
${\cal R}_{abcd}$ denotes the worldvolume Riemann tensor, and $K = g^{ab} K_{ab}$,
and $G_{ab}$ denotes the worldvolume Einstein tensor. Note that for even $n$, we 
can recognize the form of the Gauss-Bonnet (GB) terms~(see \ref{eq:lag0}-\ref{eq:lag4}) 
but expressed now in terms of the worldvolume extrinsic curvature. For example, for 
$n=0$ we have the Dirac-Nambu-Goto (DNG) Lagrangian, for $n=2$ we have the Ricci scalar Lagrangian 
also known as Regge-Teitelboim model \cite{rt,davidson,ostrogradski} and for $n=4$ 
we have the GB Lagrangian which for $p>3$ produces non-vanishing equations of motion 
with ghost-free contribution; in fact, for $p=3$ the GB combination is a total 
divergence  and it is a topological invariant. On the other hand, for odd $n$, the 
Lagrangians are seen as the Gibbons-Hawking-York boundary terms which may exist if 
we have the presence of bulk Lovelock invariants (see \ref{app:2}). 

Some remarks are in order. To avoid ambiguities for the possible gauge invariance for 
the case of odd $n$ Lagrangians, it is necessary to make a choice in the direction 
for the normal vector to $\Sigma$ in order to have a theory defined on the right-hand 
side of the worldvolume; hence, we assume that $n^\mu$ is such that it is 
pointing outward of the region of interest. In addition, we point out that the 
Lovelock brane invariants have a linear dependence on the acceleration of the brane. This remarkable 
fact leads to explore alternative Hamiltonian constructions since, as happens in 
some braneworld scenarios, when one tries to study the symmetries or quantize canonically, 
we cannot obtain quadratic constraints in the 
momenta in a straightforward way~\cite{rt,davidson,ostrogradski,paul}.

%%%%%%%%%%%%%%%%%%%%%%%%%%%%%%%%%%%%%%%%%%%%%%%%%%%%%%%%%%%%%%%%%%

\subsection{Lovelock brane tensors and equations of motion}

%%%%%%%%%%%%%%%%%%%%%%%%%%%%%%%%%%%%%%%%%%%%%%%%%%%%%%%%%%%%%%%%%%%%%

By virtue of the properties of the gKd function one can define the brane-conserved 
tensors as follows
\begin{equation}
\label{eq:conserved1}
J^{a}_{(n)b} := \delta^{a a_1 a_2 a_3 \ldots a_n} _{b b_1 b_2 b_3 \ldots b_n}
K^{b_1}{}_{a_1} K^{b_2}{}_{a_2} K^{b_3}{}_{a_3} \ldots K^{b_n}{}_{a_n}.
\end{equation}
They are symmetric and obey that $\nabla_a J^{ab} _{(n)} = 0$. This fact 
is shown by using the properties of the gKd and the Codazzi-Mainardi 
integrability condition in a flat background spacetime, $\nabla_a K_{bc} = \nabla_b 
K_{ac}$. Note that for a $(p+1)$-dimensional $\Sigma$ there are at most an equal 
number of conserved tensors $J^{ab}$. 
As developed above, by expanding out the determinant in~(\ref{eq:conserved1}) 
in terms of minors, we obtain a recursion relation\footnote{Relation 
(\ref{eq:identity}) was introduced by mathematicians under the name of 
Newton transformation \cite{reilly73}. In fact, the framework given 
in~(\ref{eq:lovelock-brane}) was outlined in \cite{reilly73,guven2} but, from 
our perspective, it lacks a physical insight.}
\begin{eqnarray}
J^a _{(n) b} 
&=& \left[ \delta^a _b \delta^{a_1 a_2 a_3 \ldots a_n} _{b_1 b_2 b_3 \ldots b_n} 
- \delta^a _{b_1} \delta^{a_1 a_2 a_3 \ldots a_n} _{b\, b_2 b_3 \ldots b_n}
+ \ldots + (-1)^n \delta^a _{b_n} \delta^{a_1 a_2 a_3 \ldots a_n} _{b \,b_1\, b_2
  \ldots b_{n-1}} \right] \times
\nonumber
\\
&& \quad \quad K^{b_1}{}_{a_1}K^{b_2} {}_{a_2} K^{b_3} {}_{a_3} 
  \ldots K^{b_n} {}_{a_n},
\nonumber
\\
&=& \delta^a _b \,L_n - n K^a{}_c J^{c}_{(n-1)b}.
\label{eq:identity}
\end{eqnarray}
The recurrent use of this identity allows us to have an expression for the 
conserved tensors
\begin{equation}
J^{ab} _{(n)} = \sum_{s=0} ^{n} \frac{(-1)^s n!}{(n - s)!}\,K^{ab}_{(s)} \,L_{n - s},
 \qquad \qquad n=0,1,2,3,4, \ldots,p
\label{eq:Jab}
\end{equation}
where we have adopted the notation: $K^{ab} _{(0)} = g^{ab}$, $K^{ab} _{(1)} = 
K^{ab}$, $K^{ab} _{(2)} = K^a{}_c K^{bc}$ and so forth. From equation (\ref{eq:Jab}),
we may compute the first conserved tensors
\numparts
\label{Jterms}
\begin{eqnarray}
J_{(0)} ^{ab} &=& g^{ab} = - 2 G_{(0)} ^{ab},
\label{eq:cons0}
\\
J_{(1)} ^{ab} &=& g^{ab} L_1 - K^{ab},
\label{eq:cons1}
\\
J_{(2)} ^{ab} &=& - 2 G_{(1)} ^{ab},
\label{eq:cons2}
\\
J_{(3)} ^{ab} &=& g^{ab}\, L_3 - 3\, {\cal R}\, K^{ab} +
 6\, K \,K^a{}_c K^{c b} - 6\, K^a{}_c K^c{}_d K^{d b}, 
\label{eq:cons3}
\\
J_{(4)} ^{ab} &=& - 2 G^{ab} _{(2)},
\label{eq:cons4}   
\\
J_{(5)} ^{ab} &=& g^{ab} \,L_5 - 5 \,L_4\, K^{ab} 
+ 20\, L_3\, K^a{}_c K^{c b} - 60\, L_2\, K^a{}_c K^c{}_d K^{d b}
\nonumber
\\
&+&  120 \,L_1 \,K^a{}_c K^c{}_d K^{d}{}_e K^{e b} 
-120\,L_0\, K^a{}_c K^c{}_d K^{d}{}_e K^{e}{}_f K^{fb},
\label{eq:cons5}
\\
J_{(6)} ^{ab} &=& - 2 G_{(3)} ^{ab},
\label{eq:cons6}   
%\\
%J_{(7)} ^{ab} &=& g^{ab}\, L_7 - 7\, L_6 \,K^{ab} 
%+ 42\, L_5\, K^a{}_c K^{cb} 
%- 210\, L_4\, K^a{}_c K^c{}_d K^{db} 
%\nonumber
%\\
%&+& 840 \,L_3\, K^a{}_c K^c{}_d K^{d}{}_e K^{eb}
%- 2520 \,L_2\,K^a{}_c K^c{}_d K^{d}{}_e K^{e}{}_f K^{fb}
%\nonumber
%\\
%&+&
%5040 \,L_1\, 
%K^a{}_c K^c{}_d K^{d}{}_e K^{e}{}_f K^{f}{}_h K^{hb}
%\nonumber
%\\
%&-& 5040\, L_0\, K^a{}_c K^c{}_d K^{d}{}_e K^{e}{}_f K^{f}{}_h 
%K^{h}{}_l K^{l b},
%\label{eq:cons7}
\end{eqnarray}
\endnumparts
where $G_{ab} ^{(n)}$ are defined in \ref{app:1}. In view of these facts, 
$J^{ab} _{(n)}$ are to be referred to as the {\it Lovelock brane tensors}. 
As a consequence of equation (\ref{eq:lovelock-brane}), the contraction of 
(\ref{eq:conserved1}) with the extrinsic curvature provides an identity 
among the Lovelock brane tensors and the Lovelock brane Lagrangians 
\begin{equation}
J^{ab} _{(n)} K_{ab} = L_{n+1}.
\label{eq:relation}
\end{equation}
It follows immediately from equations (\ref{eq:identity}) and (\ref{eq:relation}) 
that $J^{a}_{(n)a} = (p+1-n)L_n= (N-n-1)L_n$.

The main fact behind the Lagrangians~(\ref{eq:lovelock-brane}) is that their associated 
equations of motion are of second order in the derivatives of the embedding functions. 
To prove this, we shall use the so-called conserved stress tensor associated to each 
term in~(\ref{eq:Lbaction}) defined as follows~\cite{noether,aux}
\begin{equation}
f^{a\,\mu }_{(n)} = \left( L_n \,g^{ab} - L_n ^{ac} K^b{}_c \right)e^\mu {}_b
+ \left( \nabla_b L_n ^{ab} \right) n^\mu,
\label{eq:fmu1}
\end{equation}
where $L_n ^{ab} := \partial L_n / \partial K_{ab}$. It is conserved in the 
sense that $\nabla_a f^{a\,\mu} _{(n)} = 0$. This geometrical object is a powerful 
tool to study the mechanical content  of branes~\cite{noether}. In our case, from equations 
(\ref{eq:lovelock-brane}) and (\ref{eq:conserved1}), we find that $L_n ^{ab} = 
n J^{ab} _{(n-1)}$. Notice that in some sense, $J_{n-1} ^{ab}$ is the derivative of 
$L_n$. Thus, we obtain
\begin{equation}
f^{a\,\mu }_{(n)} %= \left( L_n \,g^{ab} - n J^{ac} _{(n-1)} K^b{}_c \right)e^\mu {}_b
=  J^{ab}_{(n)}\, e^\mu {}_b,
\label{eq:fmu2}
\end{equation}
where we have considered the conservation property of $J^{ab} _{(n)}$ as well as 
identity~(\ref{eq:identity}). Note further that, $f^{a\,\mu }_{(n)}$ is only 
tangential to $\Sigma$. 
%This fact is related to the appearance of second-order 
%equations of motion. 
On the physical grounds, (\ref{eq:fmu2}) is merely the linear momentum density of the 
brane whose dynamics is governed by the action 
(\ref{eq:Lbaction}) which is invariant under Poincar\'e transformations in the bulk. Following~\cite{noether}, the vanishing of the normal projection of the conservation 
law, $n\cdot \nabla_a f^{a} _{(n)} = 0$, yields the equations of motion for the 
Lagrangian~(\ref{eq:lovelock-brane}):
\begin{equation}
J^{ab} _{(n)} K_{ab} = L_{n+1} = 0,
\label{eq:eom1}
\end{equation}
whereas the tangential projection, $e_{a}\cdot \nabla_b f^{b} _{(n)} = 0$,  results 
in a geometrical identity being the conservation of 
the tensors (\ref{eq:conserved1}). Obviously, we have only one equation of 
motion which is of second order in the field variables. This fact, in 
particular, means that we have only one physical degree of freedom for this 
type of branes. 

By virtue of the definition of the extrinsic curvature $K_{ab} = - n \cdot 
\nabla_a e_b$, we can observe that the equation of motion (\ref{eq:eom1}) may also
be written as a set of conservation laws:
\begin{equation}
\label{eq:eom2}
\nabla_a \left( J^{ab} _{(n)}\,e^\mu {}_b \right) = 0,
\end{equation}
where we identify the linear momentum density (\ref{eq:fmu2}) as a 
conserved current. It should be pointed out that equations (\ref{eq:eom1}) 
and (\ref{eq:eom2}) resemble the constraints that certain higher 
co-dimensional self-gravitating branes in the Lovelock gravity obey, where, 
for instance, the Ricci scalar plays the role of correction to the DNG 
branes (see, for example, \cite{kofinas,gregory}).

%%%%%%%%%%%%%%%%%%%%%%%%%%%%%%%%%%%%%%%%%%%%%%%%%%%%%%%%%%%%%%%%%%%%%%%%%%%

\section{Born-Infeld extension of Lovelock brane gravity}

%%%%%%%%%%%%%%%%%%%%%%%%%%%%%%%%%%%%%%%%%%%%%%%%%%%%%%%%%%%%%%%%%%%%%%%%%%

Let us consider the dynamical evolution of a $p$-dimensional brane propagating
in a Minkoswki spacetime of dimension $N=p+2$ that follows from the 
local BI-type action:
\begin{equation}
\label{eq:action21}
S[X] = \Lambda \int d^{p + 1}\xi \,\sqrt{- \mathrm{det} (g_{ab} + X_{ab}
)}, 
\end{equation}
where 
\begin{equation}
X_{ab} = 2 \alpha\,K_{ab} + \alpha^2\,K_{a}{}^{c}K_{cb},
\label{eq:Xab}
\end{equation}
and $\Lambda$ is a constant with dimensions $[L]^{-p -1}$ and $\alpha$ is 
a concomitant constant with dimensions $[L]$ characterizing the relative weight 
of the nonlinear terms in the model. At first glance, this second-order action 
in the embedding variables seems to lead to fourth-order equations of motion, but this 
appearance is however deceptive. Note that $X_{ab}= X_{ba}$, contrary to the 
ordinary Dirac-Born-Infeld theory where in such a case $X_{ab} = F_{ab}$ is the 
electromagnetic field strenght\footnote{In this BI spirit, a closed approach for 
strings with $X_{ab} \propto K_a{}^{ci}K_{bci}$ was developed in \cite{lindstrom} 
where $i$ keeps the track of the number of normal vectors to the worldvolume.}. 
The action (\ref{eq:action21}) is invariant under reparameterizations of the worldvolume. 
It is expected that for the small values of $X_{ab}$ the action~(\ref{eq:action21}) 
will reproduce small correction terms to the DNG model as we will uncover later on.

This effective theory has a natural geometric interpretation. Indeed, if we consider 
the embedding
\begin{equation}
x^\mu = Y^\mu (\xi^a) = X^\mu + \alpha \,n^\mu,
\label{eq:newembedding}
\end{equation}
which is anchored to the former embedding $X^\mu$, then the tangent vectors 
are $E^\mu{}_a = e^\mu {}_a + \alpha K_a{}^b e^\mu{}_b$ where we have used the 
Gauss-Weingarten equation $\nabla_a n^\mu = K_a{}^b\,e^\mu{}_b$. The corresponding 
induced metric $M_{ab} := \eta_{\mu \nu} E^\mu{}_a E^\nu{}_b = g_{ab} + X_{ab}$ 
leads to the volume element form in (\ref{eq:action21}). The embedding 
(\ref{eq:newembedding}) is equivalent to foliate the background spacetime by 
timelike leaves along the transverse deformations of the worldvolume~\cite{rham1,trodden2,trodden}. 

The BI-type volume element form may be written in terms of the Lovelock 
brane Lagrangians for a given dimension $p$. The key to show this fact is 
to note that $M_{ab}= g_{ac} (g^c{}_d + \alpha K^c{}_d)(g^d{}_b + \alpha K^d{}_b)$. 
This entails that the action~(\ref{eq:action21}) can be rewritten as
\begin{equation}
S[X] = \Lambda \int d^{p+1}\xi \sqrt{-g}\left[ \textrm{det} \left( g^a{}_b + 
\alpha K^a{}_b \right)\right].  
\label{eq:action23}
\end{equation}
Now we turn to expand the characteristic polynomial of $K^a{}_b$ by using 
equation (\ref{eq:det-2}). It therefore follows, from equations (\ref{eq:X1}-\ref{eq:X4}) 
specialized to  $f_{ab} = \alpha K_{ab}$, that 
\numparts
\begin{eqnarray}
f_{(1)}&=& 
\alpha K, 
\label{eq:f1}
\\
f_{(2)}&=& 
\frac{\alpha^2}{2}{\cal R},
\label{eq:f2}
\\
f_{(3)}&=& 
\frac{\alpha^3}{6} \left( -K_{ab} ^3 + \frac{3}{2}K_{ab} ^2 - \frac{1}{2}
K^3 - 3 G_{ab} ^{(1)} K^{ab} \right),
\label{eq:f3}
\\
 f_{(4)}&=& \frac{\alpha^4}{24} \left( {\cal R}^2 - 4{\cal R}_{ab} 
{\cal R}^{ab} + {\cal R}_{abcd}{\cal R}^{abcd} \right),
\label{eq:f4}
\\
& \vdots &
\nonumber
\\
f_{(s)} &=& \frac{\alpha^s}{s!}\,L_s,  
\end{eqnarray}
\endnumparts
where $L_s$ is given by (\ref{eq:lovelock-brane}). Now, if we set $f_{(0)} = 1$
in equation (\ref{eq:det-2}), we may therefore write the action (\ref{eq:action23}) 
in the form
\begin{equation}
\label{eq:action22}
S[X] 
%&=& \Lambda \int d^{p+1} \xi \, \sqrt{-g}\,\left\lbrace L_0 + \alpha \,L_1
%+ \frac{\alpha^2}{2} \,L_2 + \frac{\alpha^3}{6}\,L_3  + \frac{\alpha^4}{24}\, L_4 
%+ \frac{\alpha^5}{120} \,L_5  \right. \nonumber
%\\
%&+& \left. \frac{\alpha^6}{720}\,L_6 %+ \frac{\alpha^7}{5040}\,L_7 
%+ \cdots \right\rbrace , \nonumber
%\\
=  \Lambda \int d^{p+1} \xi \, \sqrt{-g}\,\sum_{n=0} ^{p+1} \left( 
\frac{\alpha^n}{n!} \right) \, L_{n},
\end{equation}
where the expressions (\ref{eq:lovelock-brane}) have been invoked. This expansion 
clearly exhibits that the action~(\ref{eq:action21}) leads to second-order
equations of motion since each one of the emerging terms does. Moreover, from 
equation (\ref{eq:newembedding}), we can infer that when the spacetime coordinates 
suffer a deviation along the direction of the normal $n^\mu$ to the
worldvolume, the ordinary DNG volume element undergoes a deformation 
becoming in another one that can be expressed as a finite series which 
involves to the Lovelock brane invariants. %From another point of view, in 
%some sense, this expansion might be seen as a collection of second-order corrections
%to the dynamics of ordinary DNG codimension 2 $p$-branes where the lowest order
%correction is described by an action linear in the extrinsic curvature. The next
%order correction term is quadratic in the extrinsic curvature and it is provided
%by the worldvolume Ricci scalar and so on.
In a similar manner, there is an analogous construction where the strategy of 
deforming a determinant results to be useful to obtain ghost-free nonlinear massive 
gravity actions~\cite{hassan}. 

The BI-like structure outlined above is a particular case of the action~(\ref{eq:Lbaction}).
As discussed extensively in \cite{zanelli} for the gravitational case, one may relate
this type of BI-like actions to a Chern-Simons limit of the Lovelock theory by
a particular choice of the $\alpha_n$'s in the Lovelock series~(\ref{eq:Lbaction}).
On the other hand, this choice for the bulk coordinates (\ref{eq:newembedding}) 
induces, for small deformations of the brane, a certain type of brane bending mode 
in connection with the so-called Galileons, i.e., scalar fields that are claimed 
to explain the cosmic acceleration, inflation and dark energy~\cite{rham1,trodden2,trodden,galileon,deser1,van,fairlie}.

The combination $X_{ab}$ is not constrained to have the form~(\ref{eq:Xab}) only 
but it seems reasonable to think of some other possible geometric choices for $X_{ab}$ 
maintaining this BI-type structure. Some of them are not so attractive due to 
their complexity and limitation, but others still underlie the interesting geometrical 
information. For example, we can choose $X_{ab} = \alpha K_{ab} + \beta K K_{ab}$. 
In such a case, by using (\ref{eq:expansion}), the cubic expansion in small curvature 
of (\ref{eq:action21}) casts out only the first four terms of the Lovelock brane 
invariants by considering $\beta = \alpha^2 / 4$ but not beyond this. In this sense, 
the action (\ref{eq:action21}) can be also considered as a minimal Born-Infeld 
extension of the Lovelock brane theory. A special concern is provided by the unitarity 
and stability of these types of BI-like brane theories, since there are many possibilities 
that one can write for them. This issue is highly non-trivial to be resolved, unless 
further constraints are impose as the particular choices that can be made for the 
parameter $\alpha$~\cite{tekin2}. We will discuss this case elsewhere.

%%%%%%%%%%%%%%%%%%%%%%%%%%%%%%%%%%%%%%%%%%%%%%%%%%%%%%%

\subsection{Weyl invariance}

%%%%%%%%%%%%%%%%%%%%%%%%%%%%%%%%%%%%%%%%%%%%%%%%%%%%%%%%%

There are some classically equivalent actions to (\ref{eq:action21})\footnote{See 
Refs.~\cite{lindstrom2,hull,roman,guendelman} for related developments for DNG extended 
objects and also for $Dp$-branes} that might play a similar role for this model, but 
 they result again in higher nonpolynomial actions. One of the most promising 
is the one that contains a scalar field as an auxiliary field which might help
to the construction of the associated canonical formalism as well as the 
study of symmetries~\cite{roman}.

Consider the modified action
\begin{equation}
S[X^\mu,h_{ab}, \phi] = \frac{\Lambda}{2} \int d^{p+1}\xi \sqrt{-h}
\left[ \phi^{1 - \frac{2}{p+1}} h^{ab} \left( g_{ab} + X_{ab} \right)  
- \phi \left( p-1 \right)  \right],
\label{eq:action-f3} 
\end{equation}
where $h_{ab}$ is an auxiliary intrinsic metric on the worldvolume, not to
be confused with the induced metric, and $h:= \textrm{det} (h_{ab})$. Also 
here, $\phi$ is an auxiliary scalar field. 
The above action, for any $p$, is invariant under the Weyl symmetry 
\begin{equation}
h_{ab}\to h_{ab} ' = e^{2\omega({\xi})} h_{ab}, \qquad \phi \to \phi' = 
e^{-(p+1)\omega({\xi})} \phi.
\end{equation}
In addition, the action~(\ref{eq:action-f3}) is classically equivalent to the action (\ref{eq:action21}) upon the
elimination of the auxiliary fields $h_{ab}$ and $\phi$ by using its
own equations of motion. To see this, we first obtain the variation of the
action~(\ref{eq:action-f3}) with respect to $h_{ab}$:
\begin{equation}
\frac{1}{2} \left[ h^{cd}(g_{cd} + X_{cd}) - \phi^{\frac{2}{p+1}}(p-1) 
\right]h_{ab} =  g_{ab} + X_{ab}.
\label{eq:h1}
\end{equation}
We contract now with the inverse metric $h^{ab}$ to obtain $h^{ab}(g_{ab} + 
X_{ab}) = (p+1)\phi^{\frac{2}{p+1}}$. Thus, the determinant of (\ref{eq:h1}) 
is given by $\mbox{det} (g_{ab} + X_{ab}) = \left( h/2^{p+1}\right)[h^{cd}(g_{cd} 
+ X_{cd}) - \phi^{\frac{2}{p+1}}(p-1) ]^{p+1}$. On the other hand, the variation 
of the action~(\ref{eq:action-f3}) with respect to $\phi$ is
\begin{equation}
h^{ab}\left( g_{ab} + X_{ab} \right) = (p+1)\phi^{\frac{2}{p+1}}.
\label{eq:h3} 
\end{equation}
When we plug back (\ref{eq:h3}) and the determinant $h$ into the action 
(\ref{eq:action-f3}), we recover the action (\ref{eq:action21}). We believe 
that the action (\ref{eq:action-f3}) deserves a closer examination as it seems 
reasonable to think that the presence of the scalar field will allow us to make 
contact with the Galileon theory. It will be considered in a forthcoming development.

%%%%%%%%%%%%%%%%%%%%%%%%%%%%%%%%%%%%%%%%%%%%%%%%%%%%%%%%%%%%%%%%%%%%%%%%%%%%%%%%%

\section{Conclusions}

%%%%%%%%%%%%%%%%%%%%%%%%%%%%%%%%%%%%%%%%%%%%%%%%%%%%%%%%%%%%%%%%%%%%%%%%%%%%%%%%

In this work, we have first reviewed the Lovelock framework for branes. 
We have derived the general equations of motion which were written in terms 
of the conserved brane tensors. We then constructed a Born-Infeld type action 
which may be written in terms of the Lovelock brane Lagrangians for a given dimension $p$. 
In particular, the action~(\ref{eq:action21}) becomes the action 
(\ref{eq:Lbaction}) where the constants $\alpha_n$ acquire a certain form in 
terms of the concomitant constant $\alpha$. We observe further that this model 
exhibits a natural geometric interpretation. Indeed, we may think of this BI-type 
action as a modification of the ordinary DNG action in the sense that it 
is proportional to the volume as measured now by using a peculiar metric 
consisting of the induced metric $g_{ab}$ modified by some terms constructed 
from the extrinsic curvature $K_{ab}$. We may include matter fields in our 
approach, like the electromagnetic interaction via $X_{ab} \to X_{ab} + F_{ab}$, 
extending our framework to a DBI-type approach for branes. We are also interested in 
the Ostrogradski-Hamiltonian development of the Lovelock brane invariants, 
since it may help to understand  the internal symmetries of the theory as 
well as to canonically quantize braneworld models that include the Lovelock brane 
invariants. This topic will be reported 
elsewhere. In addition, a classically equivalent action to (\ref{eq:action21}) 
was provided and we analyzed its Weyl invariance through the introduction of
auxiliary fields. We hope that this BI-type effective theory may lead us to an 
entirely new geometrical standpoint for the understanding of the mechanics of
the extended objects.

\bigskip

%%%%%%%%%%%%%%%%%%%%%%%%%%%%%%%%%%%%%%%%%%%%%%%%%%%%%%%%%%%%%%%%%%%%%%%%%

\ack
We thank R. Capovilla, E. Ay\'on-Beato and A. Molgado for useful discussions 
and encouragment. We specially thank to Jemal Guven and Rub\'en Cordero for many 
enlightening discussions over several issues on the geometry of surfaces. 
MC acknowledges support from a CONACyT scholarship. ER also thanks PROMEP, 
CA-UV: Algebra, Geometr\'\i a y Gravitaci\'on. This work was partially 
supported by SNI (M\'exico).

%%%%%%%%%%%%%%%%%%%%%%%%%%%%%%%%%%%%%%%%%%%%%%%%%%%%%%%%%%%%%%%%%%%%%%%%%%

\appendix

%%%%%%%%%%%%%%%%%%%%%%%%%%%%%%%%%%%%%%%%%%%%%%%%%%%%%

\section{Lovelock Gravity Review}
\label{app:1}
%%%%%%%%%%%%%%%%%%%%%%%%%%%%%%%%%%%%%%%%%%%%%%%%%%%%%%

We give an overview of the Lovelock gravity theory. This 
is the most general theory of gravity satisfying the following three 
conditions: (1) The field equations are written in terms of a symmetric 
rank-2 tensor. (2) The theory is consistent  with the conservation law of 
the energy-momentum tensor. (3) The theory does not include higher than 
third  order derivatives of the metric. Consider an $N$-dimensional 
spacetime manifold ${\cal M}$ with metric $g_{\mu \nu}$ ($\mu,\nu=0,1,2,
\ldots,N-1$). In this spacetime, the most general action which maintains
the field equations of motion for the metric of second order, as the 
Einstein-Hilbert action, is the Lovelock gravity action \cite{lovelock} 
given by
\begin{equation}
 S[g_{\mu \nu}] = \frac{1}{2\kappa_n ^2}\int_{\cal M} d^N x\,
\sqrt{-g} \,\sum_{n=0} ^p a_n {\cal L}_n (g_{\mu \nu}, 
{\cal R}_{\mu \nu \alpha \beta}),
\label{eq:love-action}
\end{equation}
where
\begin{eqnarray}
{\cal L}_n (g_{\mu\nu}, {R}_{\mu \nu \alpha \beta}) 
&=& 
\frac{1}{2^n} \delta^{\alpha_1 \beta_1 \alpha_2 \beta_2 \ldots 
\alpha_n \beta_n} _{\mu_1 \nu_1 \mu_2 \nu_2 \ldots \mu_n \nu_n} 
\prod_{r=1} ^{n} {R}^{\mu_r \nu_r}{}_{\alpha_r \beta_r},
\end{eqnarray}
and $R_{\mu \nu \alpha \beta}$ is the spacetime Riemann tensor. Here $p:= 
\left[ \frac{N}{2} \right]$ represents the integer part of $\frac{N}{2}$, 
$a_n$ and $ \kappa_n ^2$ are the constant values. In addition, we have 
used the generalized Kronecker delta function defined by~\cite{lovelock-rund}
\begin{eqnarray}
\delta^{\alpha_1 \beta_1 \alpha_2 \beta_2 \ldots \alpha_n 
\beta_n} _{\mu_1 \nu_1 \mu_2 \nu_2 \ldots \mu_n \nu_n}
= 
\delta^{\alpha_1 \beta_1 \alpha_2 \beta_2 \ldots 
\alpha_n \beta_n} _{[\mu_1 \nu_1 \mu_2 \nu_2 \ldots \mu_n \nu_n]}
= 
\left|
\matrix{
\delta^{\alpha_1} _{\mu_1} & \delta^{\alpha_1} _{\nu_1} & \ldots & 
\delta^{\alpha_1} _{\mu_n} & \delta^{\alpha_1}_{\nu_n} 
\cr
\delta^{\beta_1} _{\mu_1} & \delta^{\beta_1} _{\nu_1} & \ldots & 
\delta^{\beta_1} _{\mu_n} & \delta^{\beta_1}_{\nu_n} 
\cr
\vdots & \vdots & \ddots & \vdots & \vdots
\cr
\delta^{\alpha_n} _{\mu_1} & \delta^{\alpha_n} _{\nu_1} & \ldots & 
\delta^{\alpha_n} _{\mu_n} & \delta^{\alpha_n}_{\nu_n} 
\cr
\delta^{\beta_n} _{\mu_{1}} & \delta^{\beta_n} _{\nu_1} & \ldots & 
\delta^{\beta_n} _{\mu_n} & \delta^{\beta_{n}} _{\nu_{n}}
}
\right|.
\end{eqnarray}
It must be noted that in $N$ dimensions, all terms for which $n> [N/2]$ are total 
derivatives and the term $n= N/2$ is the Euler density \cite{lovelock,zanelli}. 
Thus, only terms for which $n < N/2$ contribute to the field equations. The first 
four Lovelock Lagrangians are given by
\begin{eqnarray}
{\cal L}_0 &=& 1,
\label{eq:lag0}
\\
{\cal L}_1 &=& {R},
\label{eq:lag1}
\\
{\cal L}_2 &=& {R}^2 - 4{R}_{\mu \nu}R^{\mu \nu} + {R}_{\mu \nu \alpha \beta} 
{R}^{\mu \nu \alpha \beta},%=:{\cal R}_{\textrm{\tiny GB}},
\label{eq:lag2}
\\
{\cal L}_3 &=& {R}^3 - 12{R} {R}_{\mu \nu}{R}^{\mu \nu} + 16 {R}_{\mu \nu}
{R}^\mu{}_\alpha {R}^{\nu \alpha} + 24 {R}_{\mu \nu} {R}_{\alpha \beta} {R}^{\mu 
\alpha \nu \beta}
\nonumber
\\
&+& 3{R} {R}_{\mu \nu \alpha \beta}{R}^{\mu \nu \alpha \beta} 
- 24 {R}_{\mu \nu} {R}^\mu{}_{\alpha \beta \sigma} 
{R}^{\nu \alpha \beta \sigma}  
+ 2 {R}_{\mu \nu \alpha \beta}{R}^{\mu 
\nu}{}_{\rho \sigma}{R}^{\alpha \beta \rho \sigma} 
\nonumber
\\
&-& 8 {R}_{\mu \nu \alpha \beta} {R}^\mu {}_\rho {} ^{\alpha}{}_\sigma 
{R}^{\nu \sigma \beta \rho}
\label{eq:lag4}
\end{eqnarray}

A variational procedure applied to the action (\ref{eq:love-action}) 
casts out the equations of motion
\begin{equation}
 {\cal G}_{\mu \nu} = \kappa_n ^2 T_{\mu \nu},
\label{eq:eomLG}
\end{equation}
with $T_{\mu \nu}$ being the energy-momentum tensor for matter fields coming from
a possible matter action $S_{\mathrm{matter}}$ appearing in (\ref{eq:love-action}), 
and $ {\cal G}_{\mu \nu} = \sum_{n=0} ^{p} a_n G_{\mu \nu} ^{(n)}$
where the so-called Lovelock tensors $G_{\mu \nu} ^{(n)}$ are defined as
\begin{equation}
\label{eq:lovelock-tensors}
 G^{(n) \,\mu}{}_\nu = -\frac{1}{2^{n+1}}\delta^{\mu \mu_1 \mu_2 \ldots \mu_n \mu_{n+1} \ldots
\mu_{2n}} _{\nu \,\nu_1 \nu_2 \ldots \nu_n \nu_{n+1} \ldots \nu_{2n}} {R}_{\mu_1 \mu_2}{}^{\nu_1 \nu_2}
\ldots {R}_{\mu_{2n - 1} \mu_{2n}}{}^{\nu_{2n -1} \nu_{2n}}.
\end{equation}
Explicitly, the first four Lovelock tensors are 
\begin{eqnarray}
G_{\mu \nu} ^{(0)} &=& - \frac{1}{2} g_{\mu \nu} = - \frac{1}{2}g_{\mu \nu}\,{\cal L}_0,
\\
G_{\mu \nu} ^{(1)} &=& {R}_{\mu \nu} - \frac{1}{2} {R}\,g_{\mu \nu}=
{R}_{\mu \nu} - \frac{1}{2}g_{\mu \nu}\,{\cal L}_1,
\\
G_{\mu \nu} ^{(2)}   &=& 2\left( {R} {R}_{\mu \nu} - 2 {R}_{\mu \alpha}{R}^\alpha{}_\nu
- 2{R}_{\mu \alpha \nu \beta} {R}^{\alpha \beta} + {R}_\mu {}^{\alpha \beta \gamma}{R}_{\nu \alpha
\beta \gamma} \right) - \frac{1}{2}g_{\mu \nu }{\cal L}_2
\\
 G_{\mu \nu} ^{(3)} &=&3 \left( {R}^2 {R}_{\mu \nu} - 4 {R}{R}_{\mu \alpha}{R}^\alpha{}_\nu
- 4{R}_{\mu \nu} {R}^{\alpha \beta}{R}_{\alpha \beta}  + 8 {R}_{\mu \alpha}{R}_{\nu \beta}
{R}^{\alpha \beta} 
\right.
\nonumber
\\
&+& \left. 8 {R}_\mu{}^\alpha {R}_{\nu \beta \alpha \gamma}{R}^{\beta \gamma} 
+  8 {R}_\nu {}^\alpha {R}_{\mu \beta \alpha \gamma}{R}^{\beta \gamma}
+ 2{R} {R}_\mu{}^{\alpha \beta \gamma} {R}_{\nu \alpha \beta \gamma}
\right.
\nonumber
\\
&+& \left. {R}_{\mu \nu}{R}^{\alpha \beta \rho \sigma} {R}_{\alpha \beta \rho \sigma} 
- 4{R}_{\mu\alpha}{R}_{\nu \beta \rho \sigma} {R}^{\alpha \beta \rho \sigma} 
- 4{R}_{\nu \alpha }{R}_{\mu \beta \rho \sigma} 
{R}^{\alpha \beta \rho \sigma} \right. \nonumber
\\
&-& \left. 4 {R}_{\mu \alpha \beta \rho } {R}_{\nu \sigma}{}^{\beta \rho} 
{R}^{\alpha \sigma}
+ 8{R}_{\mu \alpha \nu \beta} {R}^{\alpha \rho \beta \sigma} 
{R}_{\rho \sigma} - 8 {R}_\mu{}^{\alpha \beta \rho}{R}_{\nu \alpha \beta}{}^\sigma
{\cal R}_{\rho \sigma} 
\right. \nonumber
\\
&+& \left. 
2 {R}_{\mu \alpha \beta \rho} {R}_\nu{}^{\alpha \sigma \lambda} 
{R}^{\beta \rho}{}_{\sigma \lambda} + 8 {R}_{\mu \alpha \beta \rho} 
{R}_{\nu \sigma}{}^{\beta \lambda} {R}_{\lambda}{}^{\alpha \rho \sigma}
- 4 {R}_{\mu \alpha \nu \beta} {R}^{\alpha \rho \sigma \lambda}{R}^\beta{}_{\rho \sigma
\lambda}  \right. \nonumber
\\
&-& \left. 4{R} {R}_{\mu \alpha \nu \beta}{R}^{\rho \sigma} 
+ 8 {\cal R}_{\mu \alpha \nu \beta } {R}^\alpha {}_\rho {R}^{\beta\rho} \right) -  
\frac{1}{2} g_{\mu \nu}\,{\cal L}_3.
\end{eqnarray}

%%%%%%%%%%%%%%%%%%%%%%%%%%%%%%%%%%%%%%%%%%%%%%%%%%%%%%%%%%

\subsection{Gibbons-Hawking-York-Myers boundary terms}
\label{app:2}

%%%%%%%%%%%%%%%%%%%%%%%%%%%%%%%%%%%%%%%%%%%%%%%%%%%%%%5

We review briefly the so-called Gibbons-Hawking-York-Myers (GHYM)
boundary terms. In order to have a well-posed variational principle in Lovelock 
gravity we must consider the appropriate surface terms given by \cite{myers,david,olea,mann}
\begin{equation}
 S_{b} = -2\kappa_n ^2 \int_{\partial {\cal M}} d^{N-1} x \sqrt{h}\,{\cal L}_b ^{(n)},
\end{equation}
where the Lagrangian surface terms may be written in a tensorial form as 
follows \cite{olea,mann}
\begin{eqnarray}
\fl
{\cal L}_b ^{(n)}  \sim 
\int_0 ^1 dt \,\,\delta^{a_1 a_2 \ldots a_{2n -1}} _{b_1 b_2
\ldots b_{2n-1}} K_{a_1}{}^{b_1} \left( {\cal R}_{a_2 a_3}{}^{b_2 b_3}
- 2 t^2 K_{a_2}{}^{b_2} K_{a_3}{} ^{b_3} \right) \times 
\nonumber
\\ 
\times \ldots \left( {\cal R}_{a_{2n -2} a_{2n-1}}{}^{b_{2n-2} b_{2n-1}}
- 2 t^2 K_{a_{2n-2}}{}^{b_{2n-2}} K_{a_{2n-1}}{} ^{b_{2n-1}} \right). 
\end{eqnarray}
Here $K_{\mu \nu}$ is the extrinsic curvature in the spacelike surface 
manifold $\partial {\cal M}$, and we have suppressed coupling constants. 
Hence, $K=g^{\mu \nu} K_{\mu \nu}$.
The first three counterterms to the Lovelock Lagrangians take the compact form: 
\begin{eqnarray}
{\cal L}_b ^{(1)} &=& K 
\label{eq:Lb1}
%\\
%&=& -2G_{\mu \nu} ^{(0)} K^{\mu \nu}, \nonumber
\\
{\cal L}_{b} ^{(2)} 
&=& -\frac{1}{3}K^3 + K K_{\mu \nu} ^2 - \frac{2}{3}K_{\mu \nu} ^3 - 
2 G^{(1)} _{\mu \nu}K^{\mu \nu}
\label{eq:Lb2}
%\\
%&=& J - 2 G^{(1)} _{\mu \nu}K^{\mu \nu}, \nonumber
\\
{\cal L}_b ^{(3)} &=& -\frac{2}{15} \left[  K^5 - 10 K^3 K_{\mu \nu} ^2
+ 20 K^2 K_{\mu \nu} ^3 + 15 K \left( K_{\mu \nu} ^2 \right)^2 - 30 K  
K_{\mu \nu} ^4 
\right.
\nonumber
\\
&-& 20  K_{\mu \nu} ^2  K_{\mu \nu} ^3 
+ \left. 24 K_{\mu \nu} ^5 \right]
- 2G_{\mu \nu} ^{(2)} K^{\mu \nu}.
\label{eq:Lb3} 
%\\
%&=& M - 2G_{\mu \nu} ^{(2)} K^{\mu \nu}, \nonumber 
\end{eqnarray}
%and we have introduced the quantities
%\begin{eqnarray}
%J &=& -\frac{1}{3}K^3 + K K_{ab} ^2 - \frac{2}{3}K_{ab} ^3,
%\label{J}
%\\
%M &=& -\frac{2}{15} \left[  K^5 - 10 K^3 K_{ab} ^2
%+ 20 K^2 K_{ab} ^3 + 15 K \left( K_{ab} ^2 \right)^2 - 30 K  K_{ab} ^4 
%\right.
%\nonumber
%\\
%&-& \left. 20 K_{ab} ^2  K_{ab} ^3 + 24 K_{ab} ^5  \right]. 
%\label{M}
%\end{eqnarray}

%%%%%%%%%%%%%%%%%%%%%%%%%%%%%%%%%%%%%%%%%%%%%%%%%%%%
%
\section{Expansion of the BI type action}
\label{appB}
%
%%%%%%%%%%%%%%%%%%%%%%%%%%%%%%%%%%%%%%%%%%%%%%%%%%%%%%

The determinant of a $n\times n$ matrix $A^a{}_b$ in terms of the gKd
is 
\begin{equation}
A:= \textrm{det} (A^a{}_b) = \frac{1}{n!}\delta^{a_1 a_2 a_3 
\cdots a_n } _{b_1 b_2 b_3 \cdots b_n} A^{b_1}{}_{a_1} A^{b_2}{}_{a_2}
A^{b_3}{}_{a_3} \cdots A^{b_{n}}{} _{a_{n}}.
\label{eq:determinant}
\end{equation}
In addition, the associated inverse matrix can be computed as follows:
\begin{equation}
\left( A^{-1}\right)^a{}_b = \frac{1}{(n-1)! A} \delta^{a a_2 a_3 
\cdots a_n } _{b b_2 b_3 \cdots b_n} A^{b_2}{}_{a_2}
A^{b_3}{}_{a_3} \cdots A^{b_{n}}{} _{a_{n}}.
\end{equation}
 
\noindent
The characteristic determinant of the matrix ${M}^a{}_b =
\delta^a{}_b + f^a{}_b$ may be expressed in the form~\cite{lovelock-rund}
\begin{eqnarray}
 \textrm{det} ({M}^a{}_b) &=& 1 + \sum_{s=1} ^{n} \frac{1}{s!}
\delta^{a_1 a_2 a_3 \cdots a_s} _{b_1 b_2 b_3 \cdots b_s} f^{b_1}{}_{a_1}
f^{b_2}{}_{a_2} f^{b_3}{}_{a_3} \cdots f^{b_s}{}_{a_s},
\nonumber
\\
&=& 1 +  \sum_{s=1} ^{n} f_{(s)},
\label{eq:det-2}
\end{eqnarray}
where $ s! f_{(s)} = \delta^{a_1 a_2 a_3 \cdots a_s} _{b_1 b_2 b_3 \cdots b_s} 
f^{b_1}{}_{a_1} f^{b_2}{}_{a_2} f^{b_3}{}_{a_3} \cdots f^{b_s}{}_{a_s}$ denotes 
the determinant of the $s$-rowed minor. These minors can be expressed in terms
of the traces of the $f^a{}_b$ matrix
\begin{eqnarray}
f_{(1)} &=& \delta^{a_1}_{b_1}f^{b_1}{} _{a_1} = f^a{}_a = \Tr (f),
\label{eq:X1}
\\
f_{(2)} &=& \frac{1}{2!}\delta^{a_1a_2} _{b_1b_2}f^{b_1}{}_{a_1} f^{b_2}{}_{a_2}
= \frac{1}{2} \left( \delta^{a_1} _{b_1} \delta^{a_2} _{b_2} -
\delta^{a_1} _{b_2} \delta^{a_2} _{b_1} \right)f^{b_1}{}_{a_1} f^{b_2}{}_{a_2}, 
\nonumber
\\
&=& \frac{1}{2} \left( f^a{}_a f^b {}_b - f^a{}_b f^b{}_a \right)
= \frac{1}{2} \left[ \Tr (f)^2  - \Tr (f^2 ) \right] ,
\label{eq:X2}
\\
f_{(3)} &=& \frac{1}{3!}\delta^{a_1a_2a_3} _{b_1b_2b_3}f^{b_1}{}_{a_1} f^{b_2}{}_{a_2}
f^{b_3}{}_{a_3} ,
\nonumber
\\
&=& \frac{1}{6}\left( \delta^{a_1} _{b_1}\delta^{a_2 a_3} _{b_2 b_3}
- \delta^{a_1} _{b_2} \delta^{a_2 a_3} _{b_1 b_3} + \delta^{a_1} _{b_3}\delta^{a_2 a_3} _{b_1 b_2}
 \right) f^{b_1}{}_{a_1} f^{b_2}{}_{a_2} f^{b_3}{}_{a_3},
\nonumber
\\
&=& \frac{1}{6} \left[ \delta^{a_1} _{b_1} \left( \delta^{a_2} _{b_2} \delta^{a_3} _{b_3} -
\delta^{a_2} _{b_3} \delta^{a_3} _{b_2} \right) - \delta^{a_1} _{b_2} \left( \delta^{a_2} _{b_1} 
\delta^{a_3} _{b_3} - \delta^{a_2} _{b_3} \delta^{a_3} _{b_1} \right) 
\right.
\nonumber
\\
&+& \left. \delta^{a_1} _{b_3} \left( \delta^{a_2} _{b_1} \delta^{a_3} _{b_2} - \delta^{a_2} _{b_2} \delta^{a_3} _{b_1} 
\right) \right]f^{b_1}{}_{a_1} f^{b_2}{}_{a_2} f^{b_3}{}_{a_3},
\nonumber
\\
&=& \frac{1}{6} \left[ f^a{}_a f^b {}_b f^c {}_c - 3 f^a{}_b f^b{}_a f^c{}_c + 2
f^a{}_b f^b{}_c f^c{}_a \right],
\nonumber
\\
&=& \frac{1}{6} \left[ \Tr (f)^3 - 3 \Tr (f^2) \Tr (f)
+ 2 \Tr (f^3)
\right],
\label{eq:X3}
\\
f_{(4)} &=& \frac{1}{4!}\delta^{a_1a_2a_3a_4} _{b_1b_2b_3b_4}f^{b_1}{}_{a_1} f^{b_2}{}_{a_2}
f^{b_3}{}_{a_3}f^{b_4}{}_{a_4},
\nonumber
\\
&=& \frac{1}{24}\left[ \delta^{a_1} _{b_1}\left( \delta^{a_2} _{b_2} \delta^{a_3 a_4} _{b_3 b_4}
- \delta^{a_2} _{b_3} \delta^{a_3 a_4} _{b_2 b_4} + \delta^{a_2} _{b_4}\delta^{a_3 a_4} _{b_2 b_3}
 \right)  
\right.
\nonumber
\\ 
&-& \left. \delta^{a_1} _{b_2}\left( \delta^{a_2} _{b_1} \delta^{a_3 a_4} _{b_3 b_4}
- \delta^{a_2} _{b_3} \delta^{a_3 a_4} _{b_1 b_4} + \delta^{a_2} _{b_4}
\delta^{a_3 a_4} _{b_1 b_3} \right)  \right. 
\nonumber
\\
&+& \left. \delta^{a_1} _{b_3} \left( \delta^{a_2} _{b_1} \delta^{a_3 a_4} _{b_2 b_4}
- \delta^{a_2} _{b_2} \delta^{a_3 a_4} _{b_1 b_4} + \delta^{a_2} _{b_4}
\delta^{a_3 a_4} _{b_1 b_2} \right) 
\right.
\nonumber
\\ 
&-& \left. \delta^{a_1} _{b_4} \left( \delta^{a_2} _{b_1} \delta^{a_3 a_4} _{b_2 b_3}
- \delta^{a_2} _{b_2} \delta^{a_3 a_4} _{b_1 b_3} + \delta^{a_2} _{b_3}
\delta^{a_3 a_4} _{b_1 b_2} \right)
 \right] f^{b_1}{}_{a_1} f^{b_2}{}_{a_2}
f^{b_3}{}_{a_3} f^{b_4}{}_{a_4},
\nonumber
\\
&=& \frac{1}{24}\left[  f^a{}_a f^b {}_b f^c {}_c f^d{}_d + 8 f^a{}_b f^b{}_c f^c{}_a
f^d{}_d - 6 f^a{}_b f^b{}_a f^c{}_c f^d{}_d 
\right.
\nonumber
\\
&+& \left. 3 f^a{}_b f^b{}_a f^c{}_d f^d{}_c
-  6 f^a{}_b f^b{}_c f^c{}_d f^d{}_a \right],
\nonumber
\\
&=&  \frac{1}{24}\left[ \Tr (f)^4 + 8 \Tr (f^3) \Tr (f) 
- 6 \Tr (f^2) \Tr (f)^2 + 3 \Tr (f^2)^2
\right.
\nonumber
\\
&-& \left. 6 \Tr (f^4) \right].
\label{eq:X4}
\end{eqnarray}

\noindent
In some cases it will be useful to obtain
the Taylor expansion of the square root of the characteristic determinant
(\ref{eq:det-2}) which may be obtained by using the well-known expansion
$(1+x)^{1/2} = 1 + \frac{1}{2} x - \frac{1}{8}x^2 + \frac{1}{16} x^3 - 
\cdots$ for $|x|\leq 1$. Hence, 
\begin{equation}
\fl
\left[ \textrm{det} \left(\delta^a{}_b + f^a{}_b \right) \right]^{1/2}
= 1 + \frac{1}{2}  \sum_{s=1} ^{n} f_{(s)} - \frac{1}{8} \left(  \sum_{s=1} ^{n} 
f_{(s)}\right)^2  + \frac{1}{16} \left(  \sum_{s=1} ^{n} 
f_{(s)}\right)^3 - \cdots .
\end{equation}
Thus, up to $O(f^6)$ we have
\begin{eqnarray}
\label{eq:expansion}
[\textrm{det} (\delta^a{}_b  + f^a{}_b)]^{1/2} &=& 1 + \frac{1}{2}\Tr (f) - 
\frac{1}{8} \left[  2 \Tr (f^2) - \Tr (f)^2 \right] 
\nonumber
\\
&+& \frac{1}{48} \left[ 8 \Tr (f^3) - 6 \Tr(f^2)
\Tr(f) +  \Tr(f)^3 \right]
\nonumber
\\
&-& \frac{1}{384} \left[ 48 \Tr (f^4) -32 \Tr(f^3) 
\Tr(f)  
\right. 
\nonumber
\\
&+& \left. 12 \Tr(f^2) \Tr(f)^2 -  12 \Tr(f^2)^2 - \Tr(f)^4 \right] 
\nonumber
\\
&+& \frac{1}{3840} \left[ 384 \Tr (f^5) - 240 \Tr(f^4) 
\Tr(f) 
\right. 
\nonumber
\\
&+& \left. 80 \Tr(f^3) \Tr(f)^2 - 20 \Tr(f^2) \Tr(f)^3 
\right.
\nonumber
\\ 
&+& \left. 60 \Tr(f) 
\Tr(f^2)^2 - 160 \Tr(f^2) \Tr(f^3) + \Tr(f)^5 \right]
\nonumber
\\
&-& 
%\frac{1}{46080} \left[ 3840 \Tr (X^6) - 2304 \Tr (X^5)
%\Tr (X) 
%\right.
%\nonumber
%\\
%&-& \left. 1440 \Tr (X^4) \Tr (X^2) + 720 \Tr (X^4) 
%\Tr (X)^2 
%\right.
%\nonumber
%\\
%&+& \left.  960 \Tr (X^3) 
%\Tr (X^2) \Tr (X) - 160 \Tr (X^3) \Tr (X)^3 \right.
%\nonumber
%\\
%&-& \left. 640 \Tr (X^3)^2 
%- 180 \Tr (X^2)^2 \Tr (X)^2 + 120 \Tr (X^2)^3 
%\nonumber 
%\right. 
%\\
%&+& \left. 30 \Tr (X^2) \Tr (X)^4 - \Tr (X)^6  \right] 
 O(f^6).
\end{eqnarray}

%%%%%%%%%%%%%%%%%%%%%%%%%%%%%%%%%%%%%%%%%%%%%%%%%%%%%%%%%%%%%%%%

\section*{References}

%%%%%%%%%%%%%%%%%%%%%%%%%%%%%%%%%%%%%%%%%%%%%%%%%%%%%%%%%%%%%%%%%%%


\begin{thebibliography}{99}


\bibitem{born} Born M 1934 \PRS (London) A \textbf{143} 410;
Born M and Infeld L 1934 \PRS (London) A \textbf{144} 425

\bibitem{leigh} Leigh R G 1989 {\it Mod. Phys. Lett. A} \textbf{4} 2767 

\bibitem{deser} Deser S and Gibbons G W 1998 \CQG {\textbf{15}} L35-L39

\bibitem{gibbons} Gibbons G 1998 \NP B \textbf{514} 603

\bibitem{schuller} Schuller F 2002 \APNY {\textbf{299}} 174

\bibitem{vollick} Vollick D N 2004 \PR D \textbf{69} 064030

\bibitem{nieto} Nieto J A 2004 \PR D \textbf{70} 044042

\bibitem{banados} Ba\~nados M and Ferreira P G 2010 \PRL \textbf{105} 011101

\bibitem{tekin} G\"{u}ll\"{u} I, \c{C}a\v{g}ri \c{S}i\c{s}man T
and Tekin B 2010 \CQG {\textbf{27}} 162001

\bibitem{lindstrom} Lindstrom U, Ro\v{c}ek M and Van Nieuwenhuizen P 1987
\PL B \textbf{199} 219 

\bibitem{bib} Rojas E 2011 {\it Int. J. Mod. Phys. D} \textrm{20} 59

\bibitem{dgp} Dvali G R, Gabadadze G and Porrati M 2000 \PL B \textbf{485} 208 

\bibitem{rham1} de Rham C and  Tolley A 2010 {\it J. Cosmol. Astropart. Phys.} JCAP05(2010)015 

\bibitem{trodden2} Goon G, Hinterbichler K and Trodden M 2011 {\it J. Cosmol. Astropart. Phys.} JCAP07(2011)017

\bibitem{noether} Arreaga G, Capovilla R and Guven J 2000 \APNY \textrm{279} 126

\bibitem{lovelock} Lovelock D 1971 \JMP {\bf 12} 498

\bibitem{rt} Regge T and Teitelboim C 1977 {\it Proc. Marcel
Grossman Meeting} (Trieste, Italy, 1975) ed. R Ruffini (Amsterdam: North-Holland, 1977) p 77

\bibitem{davidson} Davidson A and Karasik D 2003 \PR D {\textrm{67}} 064012

\bibitem{ostrogradski} Cordero R, Molgado A and Rojas E 2009 \PR D {\textrm{79}} 024024;
Cordero R, Molgado A and Rojas E 2011 \CQG \textbf{28} 065010

\bibitem{paul} Paul B 2013 \PR D \textbf{87} 045003

\bibitem{reilly73} Reilly R C 1973 {\it Michigan Math. J.} \textbf{20} 373

\bibitem{guven2} Guven J {\it Symmetric curvature invariants}. Unpublished
notes.

\bibitem{aux} Guven J 2004 \JPA {\bf 37} L313

\bibitem{kofinas} Charmousis C, Kofinas G and Papazoglou A 2010 {\it J. Cosmol. 
Astropart. Phys.} 01 022 

\bibitem{gregory} Bostock P, Gregory R, Navarro I and Santiago J 2004
\PRL {\bf 92} 221601

\bibitem{trodden} Goon G, Hinterbichler K and Trodden M 2011 \PRL {\textbf{106}} 231102

\bibitem{hassan} Hassan S F and Rosen R A 2011 {\it J. High Energy Phys.} 
JHEP1107(2011)009

\bibitem{zanelli} Zanelli J 2012 \CQG {\bf 29} 133001

\bibitem{galileon} Nicolis A, Ratazzi R and Trincherini E 2009 \PR D {\textbf{79}} 
064036

\bibitem{deser1} Deffayet C, Deser S and Esposito-Far\`ese G 2009 \PR D \textbf{80} 
064015;
Deffayet C, Esposito-Far\`ese G and Vikman A 2009 \PR D \textbf{79} 084003;
Deffayet C, Gao X, Steer D A and Zahariade G 2011 \PR D \textbf{84} 064039;
Deffayet C, Deser S and Esposito-Far\`ese G 2010 \PR D \textbf{82} 061501

\bibitem{van} Van Acoleyen K and Van Doorsselaere J 2011 \PR D \textbf{83} 084025

\bibitem{fairlie} Fairlie D B, Govaerts J and Morozov A 1992 \NP B \textbf{373} 214;
Fairlie D B and Govaerts J 1992 \JMP \textbf{33} 3543; Curtright T and Fairlie D {\it
A Galileon primer} \textit{arXiv:} 1212.6972 [hep-th]


\bibitem{tekin2}  G\"{u}ll\"{u} I, \c{C}a\v{g}ri \c{S}i\c{s}man T
and Tekin B 2010 \PR D {\textbf{82}} 124023


\bibitem{lindstrom2} Lindstrom U 1988 {\it Int. J. Mod. Phys. A} {\bf 3} 2401

\bibitem{hull} Abou Zeid M and Hull C 1997 \PL B \textbf{404} 264; 1998
\PL B \textbf{428} 277

\bibitem{roman} Garc\'\i a J A, Linares R and Vergara J D 2001 \PL B 
\textbf{503} 154 

\bibitem{guendelman} Guendelman E, Kaganovich A, Nissimov E
and Pacheva S 2006 \PR D \textbf{72} 086011

\bibitem{lovelock-rund} Lovelock D and Rund H 1989 {\it Tensors, 
differential forms, and variational principles} (Dover Publications, 
Inc., New York) 

\bibitem{myers} Myers R C 1987 \PR D {\textbf 36} 392

\bibitem{david} Davis S C 2003 \PR D {\textbf 67} 024030

\bibitem{olea} Olea R 2005 {\it J. High Energy Phys.} JHEP0506(2005)023

\bibitem{mann} Dehghani M H and Mann R B 2006 \PR D {\textbf 73} 104003
 
\end{thebibliography}
\end{document}